\documentstyle[prd,aps,eqsecnum,preprint,tighten]{revtex}

\def\jmp{{\it J. Math. Phys.}\ }
\def\pr{{\it Phys. Rev.}\ }
\def\prl{{\it Phys. Rev. Lett.}\ }
\def\pl{{\it Phys. Lett.}\ }
\def\np{{\it Nucl. Phys.}\ }

\def\ijmp{{\it Int. Journ. Mod. Phys.}\ }

\def\cqg{{\it Class. Quantum Grav.}\ }

\def\grg{{\it Gen. Relativ. Grav.}\ }

\def\apj{{\it Ap. J.}\ }
\def\aa{{\it Astron. Astrophys.}\ }
\def\ncim{{\it Nuovo Cim.}\ }

\def\rmp{{\it Rev. Mod. Phys.}\ }

\begin{document}
\renewcommand{\theequation}{\thesection.\arabic{equation}}
\begin{titlepage}
        \title{Curvature Quintessence}
\author{Salvatore Capozziello\thanks{E-mail: capozziello@sa.infn.it}\\
 {\em Dipartimento di Fisica "E.R. Caianiello"} \\
 {\em Istituto Nazionale di Fisica Nucleare, Sez. di Napoli,} \\
 {\em Universit\'a di Salerno, 84081 Baronissi (Sa), Italy.} \\ }
\date{\today}
\maketitle

\begin{abstract}
The issues of quintessence and cosmic acceleration can be
discussed in the framework of higher order theories of gravity.
We can define effective pressure and energy density directly
connected to the Ricci scalar of curvature of a generic fourth
order theory and then ask for the conditions to get an accelerated
expansion. Exact accelerated expanding solutions can be achieved
for several fourth order theories so that we get an alternative
scheme to the standard quintessence scalar field, minimally
coupled to gravity, usually adopted. We discuss also  conformal
transformations in order to see the links of quintessence between
the Jordan and Einstein frames.
\end{abstract}

\thispagestyle{empty} \vspace{20.mm}
 PACS number(s): 98.80.Cq, 98.80. Hw, 04.20.Jb, 04.50 \\

\vspace{5.mm}

\vfill

\end{titlepage}

\section{\normalsize\bf Introduction}

One of the astonishing recent result in cosmology is the fact that
the universe is accelerating instead of decelerating along the
scheme of standard Friedmann  model as everyone has learned in
textbooks. Type Ia supernovae (SNe Ia) allow to determine
cosmological parameters probing the today values of the Hubble
constant $H_0$ and the deceleration parameter $q_0$
\cite{perlmutter}. Besides, data coming from clusters of galaxies
at low red shift  (including the mass - to - light methods,
baryon fraction and abundance evolution) \cite{cluster}, and data
coming from the CMBR investigation (e.g. BOOMERANG)\cite{boomerang}
give observational constraints from which
we deduce the picture of a spatially flat, low density universe
dominated by some kind of non-clustered dark energy. Such an
energy, which is supposed to have dynamics, should be the origin
of the cosmic acceleration.

In terms of density parameter, we have
\begin{equation}
\label{1} \Omega_{(matter)}\simeq 0.3\,,\qquad
\Omega_{\Lambda}\simeq 0.7\,, \qquad \Omega_{k}\simeq 0.0\,\,
\end{equation}
where the $matter$ is the non-relativistic baryonic and
non-baryonic (dark) matter, $\Lambda$ is the dark energy
(cosmological constant, quintessence,..), $k$ is the curvature
parameter of Friedmann-Robertson-Walker (FRW) metric of the form
\begin{equation}
ds^2=dt^2-a(t)^2\left[\frac{dr^2}{1-k r^2}+r^2d\Omega^2\right]\,,
\end{equation}
where $a(t)$ is the scale factor of the universe.

The luminosity distance can be deduced from SNe Ia used as
standard candles.

For $z\leq 1$, by the luminosity distance $d_{L}\simeq
H_0^{-1}[z+(1-q_0)z^2/2]$, the observational results indicate
\begin{equation}
-1\leq q_0 < 0
\end{equation}
which is a clear indication for the acceleration.

The deceleration parameter can be given in terms of density
parameter and then we have for FRW models
\begin{equation}
\label{2}
q_0=-\frac{\ddot{a}a}{\dot{a}^2}=\frac{1}{2}(3\gamma+1)\Omega_{(matter)}-\Omega_{\Lambda}\,,
\end{equation}
where $\gamma$ is the constant of state equation $p=\gamma\rho$.
Immediately, we realize that acceleration or deceleration depends
on the value of $\gamma$. For standard fluid matter, it is defined
the Zeldovich range $0\leq \gamma_{(matter)} \leq 1$ where
$\gamma_{(matter)}=0$ indicates
 dust (i.e. non-relativistic matter), $\gamma_{(matter)}=1/3$ radiation
(i.e. relativistic matter). Another interesting case has been
widely considered in literature, it is $\gamma=-1$ which points
out a scalar field fluid where dynamics is dominated by
self-interaction potential or cosmological constant.  Inserting a
standard matter fluid into Friedmann--Einstein cosmological
equations gives rise to decelerated dynamics. Due to this fact,
non-standard forms of matter--energy have to be taken into account
if one wants to insert observations into a theoretical frame.

Several approaches can be pursued in order to realize  this goal.
All of them can be summarized into three great families: the
cosmological constant, the variable cosmological constant and
quintessence. Essentially they are linked but, in order to match
the observations, several issue have to be satisfied. Below we
give a short summary of this three pictures.

The cosmological constant has become one of the main issue of
modern physics since by fixing its value  should provide the
gravity vacuum state \cite{weinberg}, should make to understand
the mechanism which led the early universe to the today observed
large scale structures \cite{guth},\cite{linde}, and to predict
what will be the fate of the whole universe (no--hair conjecture)
\cite{hoyle}.

From the  cosmological point of view, the main feature of
inflationary models is the presence of a finite period during
which the expansion is de Sitter (or quasi--de Sitter or power
law): this fact implies that the expansion of the scale factor
$a(t)$ is superluminal (at least $a(t)\sim t$, in general
$a(t)\sim \exp H_{0}t$ where $H_{0}$ is the Hubble parameter
nearly constant for a finite period) with respect to the comoving
proper time $t$. Such a situation arises in  presence of an
effective energy--momentum tensor which is approximately
proportional (for a certain time) to the metric tensor and takes
place in various gravitational theories: i.e. the Einstein gravity
minimally coupled with a scalar field \cite{guth},\cite{linde},
fourth or higher--order  gravity
 \cite{starobinsky,ottewill,schmidt,kluske}, scalar--tensor
 gravity \cite{la,cimento}.

Several inflationary models are affected by the shortcoming of
"fine tuning" \cite{albrecht}, that is inflationary phase proceeds
from very special initial conditions, while a natural issue would
be to get inflationary solutions as attractors for a large set of
initial conditions. Furthermore, the same situation should be
achieved also in the future: if a remnant of cosmological
constant is today observed, the universe should evolve toward a
final de Sitter stage. A more precise formulation of such a
conjecture is possible for a restricted class of cosmological
models, as discussed in \cite{wald}. We have to note that the
conjecture holds when any ordinary matter field, satisfies the
dominant and strong energy conditions \cite{hawking}. However it
is possible to find models which explicitly violate such
conditions but satisfies no--hair theorem requests. Precisely,
this fact happens if extended gravity theories are involved and
matter is in the form of scalar fields, besides the ordinary
perfect fluid matter \cite{quartic}.

In any case, we need a time variation of cosmological constant to
get successful inflationary models, to be in agreement with
observations, and to obtain a  de Sitter stage
 toward the future. In other words, this means that cosmological
constant has to acquire a great value in early epochs (de Sitter
stage), has to undergo a phase transition with a graceful exit and
has to result in a small remnant toward the future
\cite{lambdastep}. The today observed  accelerated cosmological
behaviour should be the result of this dynamical process where the
value of cosmological constant is not fixed exactly at  zero.

In this context, a fundamental issue is to select the classes of
gravitational theories and the conditions which "naturally" allow
to recover an effective time--dependent cosmological constant
without considering special initial data.

The third approach is quintessence \cite{steinhardt}. Quintessence is a
time-varying, spatially inhomogeneous component of cosmic density
with negative pressure $-1\leq \gamma_Q\leq 0$. Formally, vacuum
energy density is quintessence in the limit $\gamma_Q\rightarrow
-1$ so that the three approaches present in literature
(cosmological constant, variable cosmological constant and
quintessence) are strictly linked.

However all of them claim for an {\it ingredient} which, a part a
pure cosmological constant, comes from a matter-energy counter
part. In this paper, we want to investigate if the quintessential
scheme can be achieved in  a geometrical way by taking into
account higher order theories of gravity.

There is no {\it a priori} reason to restrict the gravitational
Lagrangian to a linear function of the Ricci scalar $R$ minimally
coupled with matter \cite{francaviglia}. Additionally, we have to
note that, recently, some authors have taken into serious
consideration the idea that there are no "exact" laws of physics
but that the Lagrangians of physical interactions are
"stochastic" functions with the property that local gauge
invariances (i.e. conservation laws) are well approximated in the
low energy limit  and physical constants can vary \cite{ottewill}.
This scheme was adopted in order to treat the quantization on
curved spacetimes and the result was that the interactions among
quantum scalar fields and background geometry or the gravitational
self--interactions yield corrective terms in the
Einstein--Hilbert Lagrangian \cite{birrell}. Futhermore, it has
been realized that such corrective terms are inescapable if we
want to obtain the effective action of quantum gravity on scales
closed to the Planck length \cite{vilkovisky}. They are
higher--order terms in curvature invariants as $R^{2}$,
$R^{\mu\nu} R_{\mu\nu}$,
$R^{\mu\nu\alpha\beta}R_{\mu\nu\alpha\beta}$, $R\Box R$, or
$R\Box^{k}R$, or nonminimally coupled terms between scalar fields
and geometry as $\phi^{2}R$. Terms of these kinds arise also in
the effective Lagrangian of strings and Kaluza--Klein theories
when the mechanism of dimensional reduction is working
\cite{veneziano}.

Besides fundamental physics motivations, all these theories have
acquired a huge interest in cosmology due to the fact that they
"naturally" exhibit inflationary behaviours and that the related
cosmological models seem very realistic
\cite{starobinsky,la}. Furthermore, it is possible to show
that, via conformal transformations, the higher--order and
nonminimally coupled terms ({\it Jordan frame}) always
corresponds to the Einstein gravity plus one or more than one
minimally coupled scalar fields ({\it Einstein frame})
\cite{teyssandier,maeda,wands,conf,gottloeber}
so that these geometric contributions can always have a "matter"
interpretation.

Quintessence can be achieved also in the framework of
higher-order theories of gravity, that is in a geometrical way.

In Sec.II, we derive the Friedmann-Einstein equations for generic
fourth-order models. Sec.III is devoted to the discussion of
conditions to obtain quintessence while exact solutions
satisfying these prescriptions are shown in Sec.IV. Conclusions
are drawn in Sec.V.

\section{\normalsize\bf Fourth-Order Gravitational Theories
and Cosmological Equations}

A generic fourth--order theory in four dimensions can be
described by the action
 \begin{equation}\label{3}
 {\cal A}=\int d^4x \sqrt{-g} \left[f(R)+{L}_{(matter)} \right]\,{,}
 \end{equation}
 where $f(R)$ is a function of Ricci scalar $R$ and ${L}_{(matter)}$
 is the standard matter Lagrangian density.
We are using physical units $8\pi G_N=c=\hbar=1$. The field
equations are

 \begin{equation}\label{4}
 f'(R)R_{\alpha\beta}-\frac{1}{2}f(R)g_{\alpha\beta}=
 f'(R)^{;\alpha\beta}(g_{\alpha\mu}g_{\beta\nu}-g_{\alpha\beta}g_{\mu\nu})+ \tilde{T}^{(matter)}_{\alpha\beta}\,,
 \end{equation}
 which can be recast in the more expressive form
 \begin{equation}\label{5}
 G_{\alpha\beta}=R_{\alpha\beta}-\frac{1}{2}g_{\alpha\beta}R=T^{(curv)}_{\alpha\beta}+T^{(matter)}_{\alpha\beta}\,,
 \end{equation}
 where
 \begin{equation}
 \label{6}
T^{(curv)}_{\alpha\beta}=\frac{1}{f'(R)}\left\{\frac{1}{2}g_{\alpha\beta}\left[f(R)-Rf'(R)\right]+
f'(R)^{;\alpha\beta}(g_{\alpha\mu}g_{\beta\nu}-g_{\alpha\beta}g_{\mu\nu})
\right\}
 \end{equation}
 and
 \begin{equation}
 \label{7}
 T^{(matter)}_{\alpha\beta}=\frac{1}{f'(R)}\tilde{T}^{(matter)}_{\alpha\beta}\,,
 \end{equation}
 is the stress-energy tensor of matter where we have taken into account
 the nontrivial coupling to geometry. The prime means the derivative with respect to $R$.

However,  if $f(R)=R+2\Lambda$, the standard second--order gravity
is recovered.  Reducing the action to a point-like, FRW one, we
have to write
 \begin{equation}\label{8}
 {\cal A}_{(curv)}=\int dt {\cal L}(a, \dot{a}; R, \dot{R})\,{,}
 \end{equation}
where dot means derivative with respect to the cosmic time. The
scale factor $a$ and the Ricci scalar $R$ are the canonical
variables. This position could seem arbitrary since $R$ depends on
$a, \dot{a}, \ddot{a}$, but it is generally used in canonical
quantization \cite{schmidt,vilenkin,lambda}. The definition of $R$
in terms of $a, \dot{a}, \ddot{a}$ introduces a constraint which
eliminates second and higher order derivatives in action
(\ref{8}), and gives a system of second order differential
equations in $\{a, R\}$. Action (\ref{8}) can be written as
 \begin{equation}\label{10}
 {\cal A}_{(curv)}=2\pi^2\int dt \left\{ a^3f(R)-\lambda\left [ R+6\left (
 \frac{\ddot{a}}{a}+\frac{\dot{a}^2}{a^2}+\frac{k}{a^2}\right)\right]\right\}\,{,}
 \end{equation}
where the Lagrange multiplier $\lambda$ is derived by varying
with respect to $R$. It is
 \begin{equation}\label{11}
 \lambda=a^3f'(R)\,{.}
 \end{equation}
 The point-like Lagrangian is then
$$
 {\cal L}={\cal L}_{(curv)}+{\cal L}_{(matter)}=a^3\left[f(R)-R
 f'(R)\right]+6a\dot{a}^2f'(R)+
$$
\begin{equation}
\label{12}
\qquad +6a^2\dot{a}\dot{R}f''(R)-6ka
 f'(R)+a^3p_{(matter)}\,,
 \end{equation}
 where we have taken into account also the fluid matter
 contribution which is, essentially, a pressure term
 \cite{quartic}.

 The Euler-Lagrange equations are
\begin{equation}
\label{13}
2\left(\frac{\ddot{a}}{a}\right)+\left(\frac{\dot{a}}{a}\right)^2+
\frac{k}{a^2}=-p_{(tot)}\,,
 \end{equation}
and
\begin{equation}
\label{14}
f''(R)\left[R+6\left(\frac{\ddot{a}}{a}+\frac{\dot{a}^2}{a}^2+\frac{k}{a^2}\right)\right]=0\,.
 \end{equation}
The dynamical system is completed by the energy condition
\begin{equation}
\label{15}
 \left(\frac{\dot{a}}{a}\right)^2+\frac{k}{a^2}=\frac{1}{3}\rho_{(tot)}\,.
\end{equation}

\section{\normalsize\bf Curvature Quintessence}

Combining Eq.(\ref{13}) and Eq.(\ref{15}), we obtain
 \begin{equation}
 \label{16}
 \left(\frac{\ddot{a}}{a}\right)=-\frac{1}{6}\left[\rho_{(tot)}+3p_{(tot)}
 \right]\,,
 \end{equation}
 where it is clear that the accelerated or decelerated behaviour
 depends on the rhs.
 However
 \begin{equation}
 \label{17}
 p_{(tot)}=p_{(curv)}+p_{(matter)}\;\;\;\;\;\rho_{(tot)}=\rho_{(curv)}+\rho_{(matter)}\,,
 \end{equation}
where we have distinguished the curvature and matter
contributions.

From the curvature-stress-energy tensor, we can define a
curvature pressure
\begin{equation}
\label{18}
p_{(curv)}=\frac{1}{f'(R)}\left\{2\left(\frac{\dot{a}}{a}\right)\dot{R}f''(R)+\ddot{R}f''(R)+\dot{R}^2f'''(R)
-\frac{1}{2}\left[f(R)-Rf'(R)\right] \right\}\,,
 \end{equation}
and a curvature density
\begin{equation}
\label{19}
\rho_{(curv)}=\frac{1}{f'(R)}\left\{\frac{1}{2}\left[f(R)-Rf'(R)\right]
-3\left(\frac{\dot{a}}{a}\right)\dot{R}f''(R) \right\}\,.
 \end{equation}

From Eq.(\ref{16}), the accelerated behaviour is achieved if
\begin{equation}
\label{20} \rho_{(tot)}+ 3p_{(tot)}< 0\,,
\end{equation}
which means
\begin{equation}
\label{21} \rho_{(curv)}> \frac{1}{3}\rho_{(tot)}\,,
\end{equation}
assuming that all matter components have non-negative pressure.

In other words, conditions to obtain acceleration  depends on the
 relation
\begin{equation}
\label{22}
 \rho_{(curv)}+3p_{(curv)}=\frac{3}{f'(R)}\left\{\dot{R}^2f'''(R)+\left(\frac{\dot{a}}{a}\right)\dot{R}f''(R)
 +\ddot{R}f''(R)-\frac{1}{3}\left[f(R)-Rf'(R)\right]\right\}\,,
 \end{equation}
 which has to be compared with matter contribution. However, it
 has to be
 \begin{equation}
 \label{23}
 \frac{p_{(curv)}}{\rho_{(curv)}}=\gamma_{(curv)}\,,
 \qquad -1\leq\gamma_{(curv)}<0\,.
 \end{equation}
 The form of $f(R)$ is the main ingredient to obtain this {\it curvature
 quintessence}.

 \section{\normalsize\bf Exact solutions}
As simple choice in order to fit the above prescriptions, we ask
for solutions of the form
 \begin{equation}
 \label{24}
 f(R)=f_0 R^n\,,\qquad
 a(t)=a_0\left(\frac{t}{t_0}\right)^{\beta}\,.
 \end{equation}
 However, the interesting cases are for $n\neq 1$ (Einstein
 gravity) and $\beta\geq 1$ (accelerated behaviour). Inserting
 Eqs.(\ref{24}) into the above dynamical system, we obtain the exact solutions
\begin{equation}
\label{25} \beta=2\,;\qquad n=-1,3/2\,;\qquad k=0\,.
\end{equation}
In both cases, the deceleration parameter is
 \begin{equation}
 \label{26}
 q_0=-\frac{1}{2}\,,
 \end{equation}
 in perfect agreement with the observational results.

The case $n=3/2$ deserves further discussion. It is interesting in
conformal transformations from Jordan frame to Einstein frame
\cite{cqgconf,magnano} since it is possible to give explicit form
of scalar field potential. In fact, if
 \begin{equation}\label{27}
 \tilde{g}_{\alpha\beta}\equiv
 f'(R)g_{\alpha\beta}\,{,}\qquad
 \varphi=\sqrt{\frac{3}{2}}\ln f'(R)\,{,}
 \end{equation}
 we have the conformal equivalence of the Lagrangians
 \begin{equation}\label{28}
 {\cal L}=\sqrt{-g}\,f_0R^{3/2}\longleftrightarrow
 \tilde{{\cal L}}=\sqrt{-\tilde{g}}\left[-\frac{\tilde{R}}{2}+
 \frac{1}{2}\nabla_{\mu}\varphi\nabla^{\mu}\varphi-V_0\exp\left(
 \sqrt{\frac{2}{3}}\varphi\right)\right]\,{,}
 \end{equation}
 in our physical units. This is the so--called Liouville field
theory and it is one of the few cases where a fourth--order
Lagrangian can be expressed, in the Einstein frame, in terms of
elementary functions under a conformal transformation. It is
possible to obtain the general cosmological solution
\cite{lambiase} which is
 \begin{equation}\label{29}
 a(t)=a_0[c_4t^4+c_3t^3+c_2t^2+c_1t+c_0]^{1/2}\,{.}
 \end{equation}

 The constants $c_i$ are combinations of the initial conditions.
Their values determine the type of cosmological evolution. For
example, $c_4\neq 0$ gives a power law inflation while, if the
regime is dominated by the linear term in $c_1$, we get a
radiation--dominated stage.

\section{\normalsize\bf Conclusions}

In this paper, we have shown that the quintessence ''paradigm"
can be recovered in the framework of higher-order theories of
gravity. In other words, as it is possible for inflationary
models, we can ask for a sort of {\it curvature quintessence}
which can be recovered by taking into account curvature geometric
invariants. The interest of this approach is that quintessence
could be related to some effective theory of quantum gravity
where curvature invariants are widely derived \cite{odintsov}.

In any case, the models have to be improved by the comparison
with observations in order to see if it is possible to constrain
the form of $f(R)$ without making {\it ad hoc} choices. In this
sense, some results are present in literature where the form of
$f(R)$ is selected by the CMBR constraint \cite{hwang}.

\section*{Acknowledgement}
The author is grateful to A.A. Starobinsky  for the useful
discussions and comments on the topics.

\end{document}